\documentclass[aps,prl,twocolumn,showpacs]{revtex4-1}

\usepackage{graphicx} \usepackage{amsmath}
\usepackage[colorlinks]{hyperref}

\newcommand{\EqLabel}[1]{\label{#1}} \newcommand{\mb}[1]{\mathbf{#1}}
\newcommand{\mr}[1]{\mathrm{#1}}

\begin{document}
 
\title{Optical conductivity of the Holstein polaron}

\author{Glen L. Goodvin$^1$, Andrey S. Mishchenko$^{2,3}$ and Mona Berciu$^1$} 

\affiliation{$^1$Department of Physics \&  Astronomy, University of
  British Columbia, Vancouver, BC, Canada, 
  V6T 1Z1\\
$^2$Cross-Correlated Materials   Research Group (CMRG), ASI, RIKEN,
  Wako 351-0198, Japan\\
$^3$Russian Research Centre ``Kurchatov Institute'',
  123182  Moscow, Russia}

\date{\today}
 
\begin{abstract} 
The Momentum Average approximation is used to derive a new kind of
non-perturbational analytical expression for the optical conductivity
(OC) of a Holstein polaron at zero temperature. This provides insight
into the shape of the OC, by linking it to the structure of the
polaron's phonon cloud.  Our method works in any dimension, properly
enforces  selection rules, can be
systematically improved, and also generalizes to momentum-dependent couplings.  Its
accuracy is demonstrated by a comparison with the first detailed set
of three-dimensional numerical OC results, obtained using the
approximation-free diagrammatic Monte Carlo method.
\end{abstract}


\maketitle


Although the study of polarons is one of the older problems in solid
state physics~\cite{landau:33}, a full understanding of their
properties is still missing.  This is especially true for the excited
states which influence response functions like the optical
conductivity (OC).  OC measurements have revealed the role of the
electron-phonon (el-ph) coupling in many materials, {\em e.g.,}
cuprates~\cite{calvani:97} and manganites \cite{Gunna:08}. In particular, the {\em
shape} of the OC curve is important, as it signifies large versus
small polaron behavior \cite{hartinger:06}.

The OC of polarons has been studied numerically using exact
 diagonalization in 1D \cite{loos:07} and for small clusters in higher
 dimension (here, finite size effects can be an
 issue)~\cite{fehske:97}.  There are also some diagrammatic Monte
 Carlo (DMC) results for the 3D Fr\"ohlich and 2D Holstein models
 \cite{mishchenko:03}.  DMC gives approximation-free results in the
 thermodynamic limit, but it requires significant computational
 effort; this is why there are very few DMC OC sets available in the
 literature.  The conceptual problem associated with all numerical
 methods, however, is that they do not provide much insight for
 understanding the shape of the OC and its relation to the properties
 of the polaron. Analytical expressions are needed for this, but most
 prior work was limited to perturbational regimes \cite{Emin}. The one
 exception is work based on the dynamical mean-field theory (DMFT)
 \cite{Fratini}, which however ignores current vertex corrections. The
 consequences are discussed below; here we state only that 
a complete understanding of the shape of the OC is still not achieved
using it.

It is, then, hard to overemphasize the need for an accurate analytical
expression establishing a nonperturbative structure of the OC.  In
this Letter we obtain such an expression using a
generalization of the Momentum Average (MA) approximation.  MA was
developed for the single-particle Green's function of the
Holstein polaron \cite{berciu:06} and then extended to more complex
models \cite{goodvin:08}, including  disorder \cite{berciu:10}. It
is non-perturbational since it sums \emph{all} 
self-energy diagrams, up to exponentially small terms which are
discarded. MA becomes exact in various asymptotic limits, satisfies
multiple spectral weight sum rules, is quantitatively accurate in
any dimension, at all energies for all parameters except in the
extreme adiabatic limit, and can be systematically
improved~\cite{berciu:06}.

Here we show how to use MA to calculate  two-particle
Green's functions, needed in response functions like the OC. Besides
efficient yet accurate results at any coupling, this finally provides
the explanation for the physical meaning of the shape of the
OC. Moreover, this MA-based approach generalizes  to OC  calculations
for models with momentum-dependent el-ph coupling  
\cite{goodvin:08}.   

We use the Holstein model~\cite{holstein:59} as a specific example since
some numerical data 
is available for comparison:
\begin{equation}
\nonumber {\cal H} = \sum_{\mb{k}} \left(
\varepsilon^{}_{\mb{k}}c^{\dagger}_{\mb{k}}c^{}_{\mb{k}} + \Omega
b^{\dagger}_{\mb{k}} b^{}_{\mb{k}} \right) + \frac{g}{\sqrt{N}}
\sum_{\mb{k}, \mb{q}} c^{\dagger}_{\mb{k} - \mb{q}}c^{}_{\mb{k}}
\left( b^{\dagger}_{\mb{q}} + b^{}_{-\mb{q}} \right).
\end{equation}
Here, $c_{\mb{k}}^{\dagger}$ and $b_{\mb{k}}^{\dagger}$ are electron and
boson creation operators for a state of momentum $\mb{k}$
(the electron's spin
is trivial and we suppress its index). The free electron dispersion
$\varepsilon_{\mb{k}} = -2t\sum_{i=1}^d \cos(k_i a)$ is for
nearest-neighbor hopping on a $d$-dimensional hypercubic lattice of
constant $a$, and the Einstein optical phonons have energy
$\Omega$.  The last term describes the local el-ph coupling $g\sum_i
c_i^{\dagger} c_i^{}(b_i^{\dagger} + b_i^{})$, written in
$\mb{k}$-space. All sums over momenta are over the Brillouin
zone and we take the total number $N$ of sites  to infinity. We set
$\hbar=1$ and $a=1$ throughout.

For the case we study here, {\em i.e.} a single polaron at $T=0$, the
optical conductivity is given by the Kubo formula \cite{mahan:81}: 
\begin{equation} \label{eq:kubo}
\sigma(\omega) = \frac{1}{\omega V} \int_0^{\infty} dt e^{i\omega t}
\langle \psi_0 | [\hat{j}^{\dagger}(t), \hat{j}(0)] | \psi_0 \rangle,
\end{equation}
with $V$ the volume, $|\psi_0 \rangle$ the
polaron ground state (GS), and the charge current operator $\hat{j}= 2
e t \sum_{\mb{q}} \sin 
q^{} c_{\mb{q}}^{\dagger}c_{\mb{q}}^{}$ is
in the Heisenberg picture. Here,  $e$ is the electron charge
and $q$ the component of $\mb{q}$ parallel to the electric
field.

Our main result is that the optical absorption equals:
\begin{equation} 
\label{eq:cond_MA}
\mr{Re} \left[\sigma_{\mr{MA}^{(i)}}(\omega)\right] = \frac{4\pi e^2
  t^2}{\omega} \sum_{n \ge 1} P_n^{(i)} f_n^{(i)}(\omega).
\end{equation}
Here $i\ge 0$ is the level of MA$^{(i)}$ approximation, denoting
an increasing complexity of the variational description of the
eigenstates~\cite{berciu:06}.  However, the  physical
meaning is the same:
$P_n^{(i)}$ is the GS probability to have $n$ phonons at the 
electron site, while $f_n^{(i)}(\omega)$ are spectral
functions describing
the electron's optical absorption in this $n$-phonon environment.
Eq. (\ref{eq:cond_MA})  shows the direct link between the OC and the
structure of the 
polaron's phonon cloud. 

Further discussion is provided below. First, we derive
Eq. (\ref{eq:cond_MA}) so that the meaning of various quantities
becomes clear.  Expanding the commutator and doing the integral in
Eq. (\ref{eq:kubo}), we find $\sigma(\omega) = \sigma_+(\omega)
+\left( \sigma_+(-\omega)\right)^*$, with:
\begin{equation}
\EqLabel{n1} \sigma_+(\omega) = {i\over \omega V} \langle \psi_0
|\hat{j}\hat{G}(\omega+E_0)\hat{j} |\psi_0\rangle,
\end{equation}
where $\hat{G}(\omega) = [ \omega + i \eta - {\cal H}]^{-1}$ with
$\eta \rightarrow 0_+$, and $E_0$ is the polaron GS energy. The usual
route is to use a Lehmann representation, leading to the well-known
formula:
\begin{equation}
\EqLabel{n2}
\sigma(\omega) ={\pi\over \omega V} \sum_{n}^{} |\langle
\psi_0|\hat{j}|\psi_n\rangle|^2 \delta(\omega+E_0-E_n)
\end{equation}
in terms of
excited polaron eigenstates $|\psi_n\rangle$,  $E_n$.
Instead, we use twice the resolution of 
identity to rewrite:
\begin{multline}
\label{simp}
\sigma_+(\omega) = \frac{i(2et)^2}{\omega V} \sum_{\mb{q},\mb{Q}}\sin q
\sin Q  \sum_{\alpha,\beta}^{} \langle \psi_0 |
c_{\mb{q}}^{\dagger} | \alpha\rangle \\ \times F_{\alpha\beta}(\mb{q},
\mb{Q},\omega+E_0)\langle \beta | 
c_{\mb{Q}}^{} | \psi_0\rangle,
\end{multline}
where $F_{\alpha\beta}(\mb{q},\mb{Q},\omega)= \langle \alpha|
c_{\mb{q}} \hat{G}(\omega) c_{\mb{Q}}^{\dagger}|\beta\rangle$.
Since $|\psi_0 \rangle$ is the polaron GS, 
$\{|\alpha\rangle\}$ and $\{|\beta\rangle\}$ are  phonon-only
states. Moreover, because of invariance to translations,
their momentum is $-\mb{q}$, respectively $-\mb{Q}$. 
Eq. (\ref{simp})  is  exact.

Consider now these matrix elements within MA$^{(0)}$, whose
 variational meaning is to expand polaron eigenstates in the basis $\{
 c_i^\dagger (b_j^\dagger)^n|0\rangle\}$, $(\forall) i, j,
 n$~\cite{berciu:06,bar:07}. Then, $ |\alpha\rangle
 \rightarrow|-\mb{q},n \rangle = \frac{1}{\sqrt{N}}\sum_i e^{-i\mb{q}
 \cdot \mb{R}_i} (b_i^{\dagger })^n | 0 \rangle $, since only such
 states will have finite overlaps in Eq. (\ref{simp}). The sums over
 $\alpha,\beta$ are now sums over phonons numbers $n, m\ge1$. Note that
 $n,m=0$ do not contribute to the regular part of $\sigma(\omega)$
 since $|-\mb{q},0 
 \rangle\sim \delta_{\mb{q},0}|0\rangle $, and $\sin q
 \cdot\delta_{\mb{q},0}\rightarrow 0$.  They do contribute to the
 Drude peak, ${\cal D}\delta(\omega)$.

The calculation of the single electron Green's functions
$F_{nm}(\mb{q},\mb{Q},\omega)= \langle -{\mb q},n| c_{\mb{q}}
\hat{G}(\omega) c_{\mb{Q}}^{\dagger}|-{\mb Q},m\rangle$ and of the
residues $\langle \psi_0 | c_{\mb{q}}^{\dagger} | -{\mb q},n\rangle$
is now carried out. Details are provided in the supplementary
material at the end of this article. Here, we note that because of the odd $\sin
q, \sin{Q}$ terms, only the part of $F_{nm}(\mb{q},\mb{Q},\omega)$
proportional to $ \delta_{\mb{q},\mb{Q}} \delta_{n,m}$ has
non-vanishing contribution to Eq.  (\ref{simp}), explaining the single
sum over $n\ge 1$ in Eq. (\ref{eq:cond_MA}). Finally, the overlap
$|\langle \psi_0 | c_{\mb{q}}^{\dagger} | -{\mb q},n\rangle|^2$ is
linked to the probability $P_n$ to have $n$ phonons at the electron
site in the polaron GS~\cite{berciu:08}. The $P_n$  expressions are listed
in the supplementary material. 
Altogether, we obtain the result of 
Eq. (\ref{eq:cond_MA}), where:
\begin{multline}
\label{fn}
 f^{(0)}_n(\omega) ={1\over
  N}\sum_{\mb{q}} \sin^2 q \left|\frac{G_0(\mb{q}, E_0 - n
  \Omega)}{\bar{g}_0(E_0 - n\Omega)} \right|^2 \\ \times
[ A_0(\mb{q}, \omega +
  E_0 - n\Omega) - A_0(\mb{q}, -\omega + E_0 - n\Omega) ],
\end{multline}
 with
$A_0(\mb{q},\omega) = -{1\over\pi} \mr{Im} \, G_0(\mb{q},\omega)$
 being the free
electron spectral weight, where
$G_0(\mb{q},\omega)=(\omega+i\eta-\varepsilon_{\mb{q}})^{-1}$. Within
the $n$-phonon sector of $|\psi_0\rangle$, the electron can
carry any momentum $\mb{q}$ and $ f^{(0)}_n(\omega)$ basically describes its
optical absorption in the presence of this  phonon environment.

At the MA$^{(1)}$ level, the variational basis is
supplemented with the states $\{ c_i^\dagger
(b_j^\dagger)^nb^\dagger_l|0\rangle\}$, $(\forall) i, j\ne l, n$, which
are key to describe the polaron+one-phonon
continuum~\cite{berciu:06}. This enlarges the 
$\{|\alpha\rangle\}$ and $\{|\beta\rangle\}$ sets with the states
$ |-\mb{q},n, \delta \rangle = \frac{1}{\sqrt{N}}\sum_i e^{-i\mb{q}
  \cdot \mb{R}_i} (b_i^{\dagger })^n b_{i+\delta}^\dagger| 0 \rangle
$ for all $\delta\ne 0$.  The new matrix elements  are found similarly and
give much lengthier yet more accurate formulae for $P_n^{(1)}$
and $f_n^{(1)}(\omega)$~\cite{goodvin:10}, but with the same physical
 meaning.

We now discuss the key features of the 3D OC results. This is because (i)
no accurate Holstein OC results were available in 3D, and we provide the first
detailed set of DMC data; and (ii) DMFT is a better
approximation for higher D. For the OC, the key approximation of DMFT
is to ignore current vertex corrections, so we test its validity in
3D where it should be most accurate.

\begin{figure}[t]
\includegraphics[width=\columnwidth]{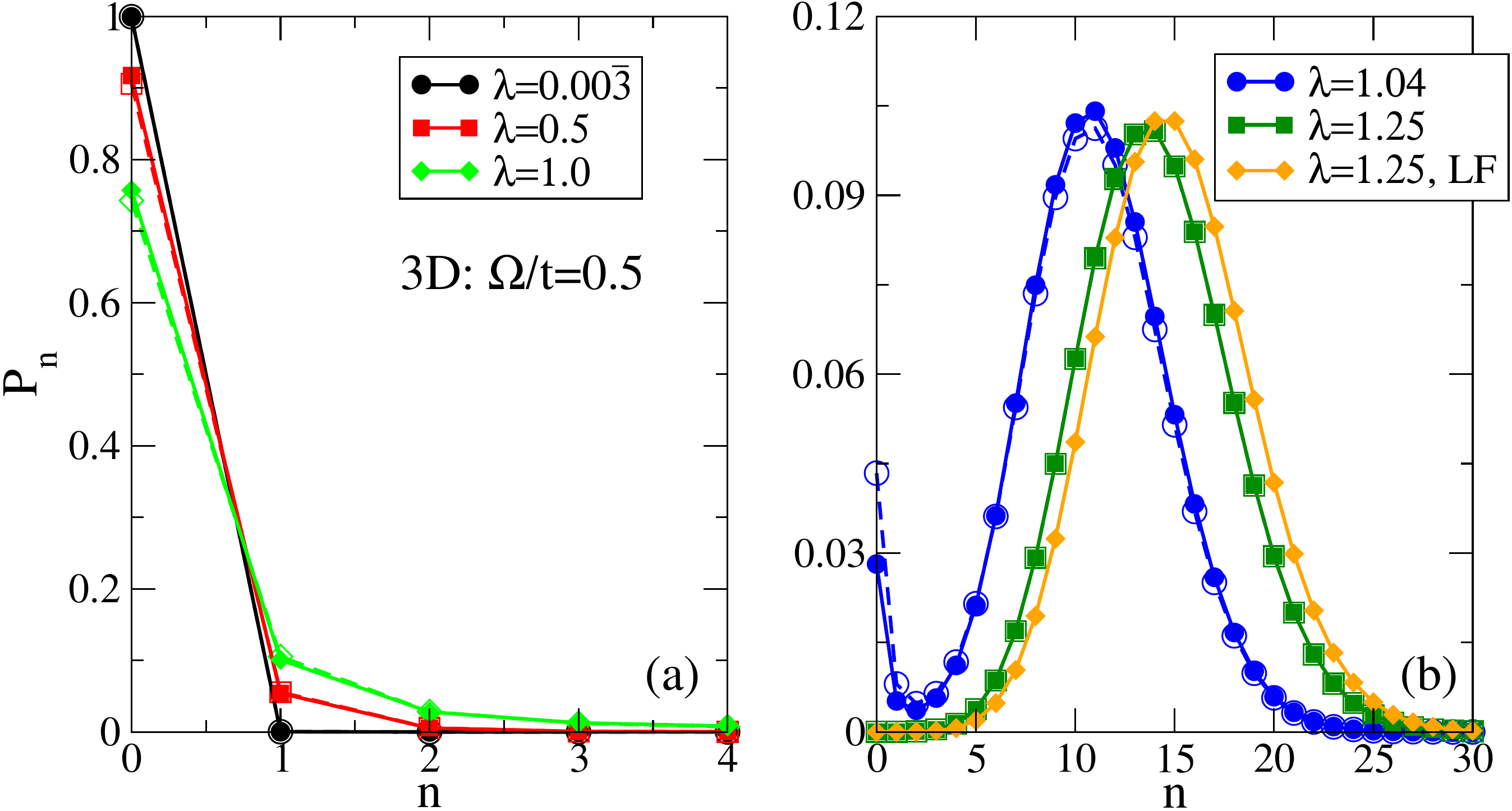}
\caption{(color online) Phonon statistics for a 3D Holstein polaron
  with (a) small to intermediate, and (b) large effective couplings,
  and $\Omega/t=0.5$.  MA$^{(0)}$ (dashed lines, open symbols) and
  MA$^{(1)}$ (solid lines, solid symbols) results are shown. The
  Lang-Firsov (LF) asymptotic result is shown for $\lambda=1.25$.
\label{fig:Pn_3D}}
\end{figure}

GS phonon probabilities $P_n$  for a 3D Holstein polaron at several effective
couplings $\lambda = g^2/(2dt \Omega)$ are shown in
Fig. \ref{fig:Pn_3D}. 1D results are quite
similar~\cite{berciu:08}. At weak el-ph couplings, the  polaron has few phonons in its
cloud so $P_{n\ge1}$ are small.  At the crossover to a small
polaron at $\lambda \sim 1$, the distribution changes abruptly and
dramatically.  As $\lambda \rightarrow \infty$, $P_n\rightarrow
\frac{1}{n!}\left( \frac{g}{\Omega} \right)^{2n} e^{-{g^2\over
    \Omega^2 }}$, the Lang-Firsov (LF) limit, already quite
accurate for $\lambda=1.25$.
 The difference between  MA$^{(0)}$ and
MA$^{(1)}$ is small. This is not surprising, since these probabilities
are for the GS, which is already very accurately described by
MA$^{(0)}$~\cite{berciu:06}. The additional basis states added at the MA$^{(1)}$
level are essential to describe excited states in the
polaron+one-phonon continuum, starting at $\Omega$ above the
GS, and due to a phonon excited \emph{far} from the
polaron cloud~\cite{berciu:06}. As we show now, they do have a significant
effect on the $f_n(\omega)$ functions and therefore on the OC onset.

\begin{figure}[t]
\includegraphics[width=\columnwidth]{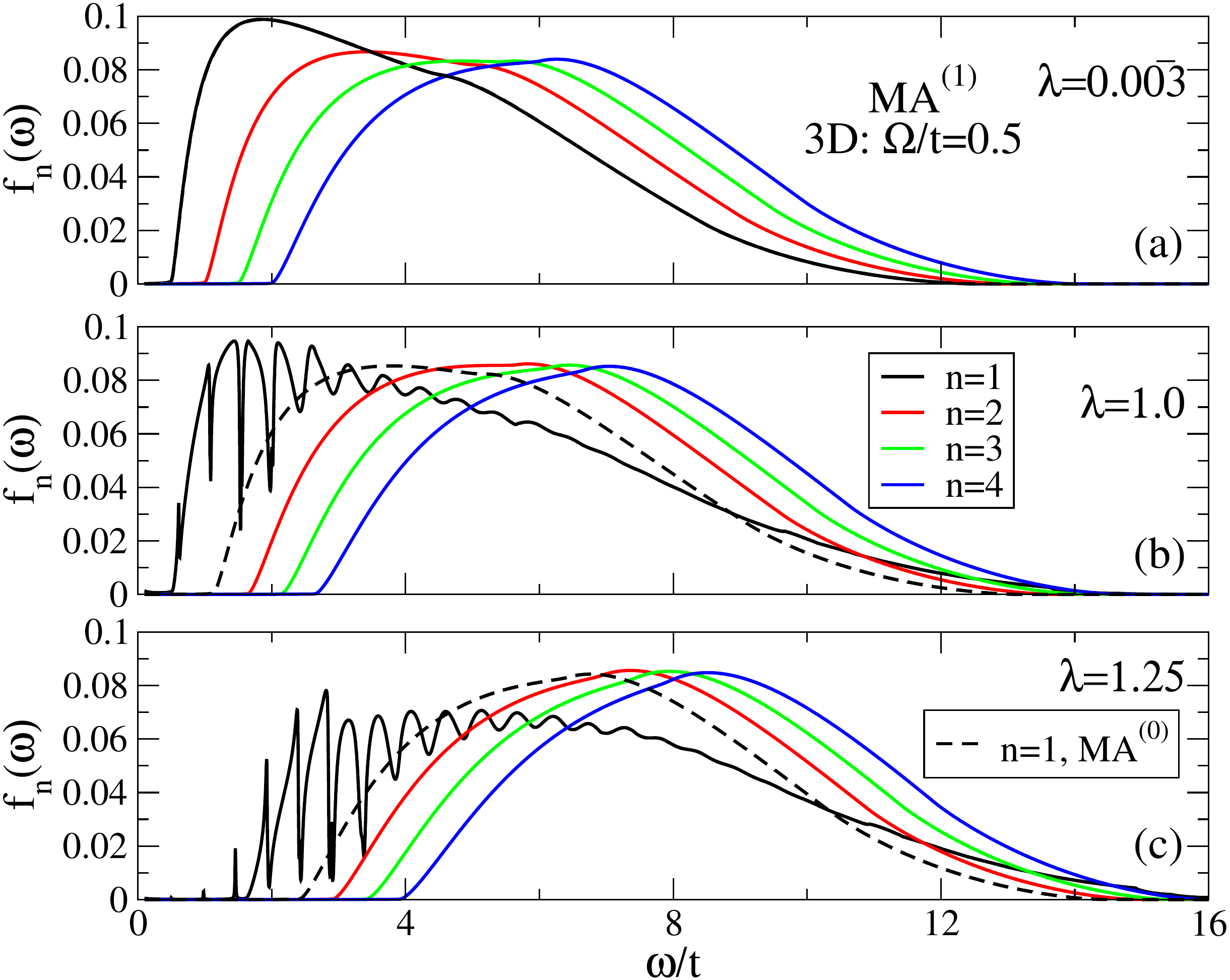}
\caption{(color online) MA$^{(1)}$ functions $f^{(1)}_n(\omega)$, for
  (a) small, (b) medium, and (c) large el-ph coupling, at
  $\Omega/t=0.5, \eta=0.005$ in 3D. Also shown is 
  $f^{(0)}_{n=1}(\omega)$ (dashed line). 
\label{fig:fn_3D}}
\end{figure}

The $\omega$-dependence of the OC is dictated by
$f_n(\omega)$. Eq. (\ref{fn}) shows that $f_n^{(0)}(\omega)$ becomes
finite at $\omega + E_0 -n\Omega \ge -2dt$, because the free-electron
spectral weight is finite in $[-2dt, 2dt]$. This implies that the
onset of absorption is set by the $n=1$ curve to be $\omega_{\rm th} = -2dt
-E_0+\Omega$, and larger $n$ contributions are shifted $(n-1)\Omega$
higher. As $\lambda$ increases and $E_0$ falls further below $-2dt$,
this suggests that $\omega_{\rm th}$ increases monotonically. This is wrong:
the OC onset is always expected at
$\omega_{\rm th}=\Omega$~\cite{loos:07,mishchenko:03}. The discrepancy is
easy to understand. The onset is due to absorption into the
polaron+one-phonon continuum, which is not described by
MA$^{(0)}$, only by MA$^{(1)}$ and higher
levels~\cite{berciu:06}. Indeed, as shown in Fig. \ref{fig:fn_3D}, 
there is a significant difference between the corresponding $n=1$
curves, and $f^{(1)}_{n=1}(\omega)$ does have an onset at
$\omega_{\rm th}=\Omega$ even though it becomes  hard to see at
larger $\lambda$. The $n\ge 2$ curves are much less affected, in
particular their onset roughly agrees with that 
predicted at MA$^{(0)}$ level. Similar behavior is found in 1D,
but the peak in each $f_n(\omega)$  moves 
towards the low-energy threshold~\cite{goodvin:10}, as expected since
the 1D free electron density of states is singular at the band-edge.

\begin{figure*}[t]
\center \includegraphics[width=2\columnwidth]{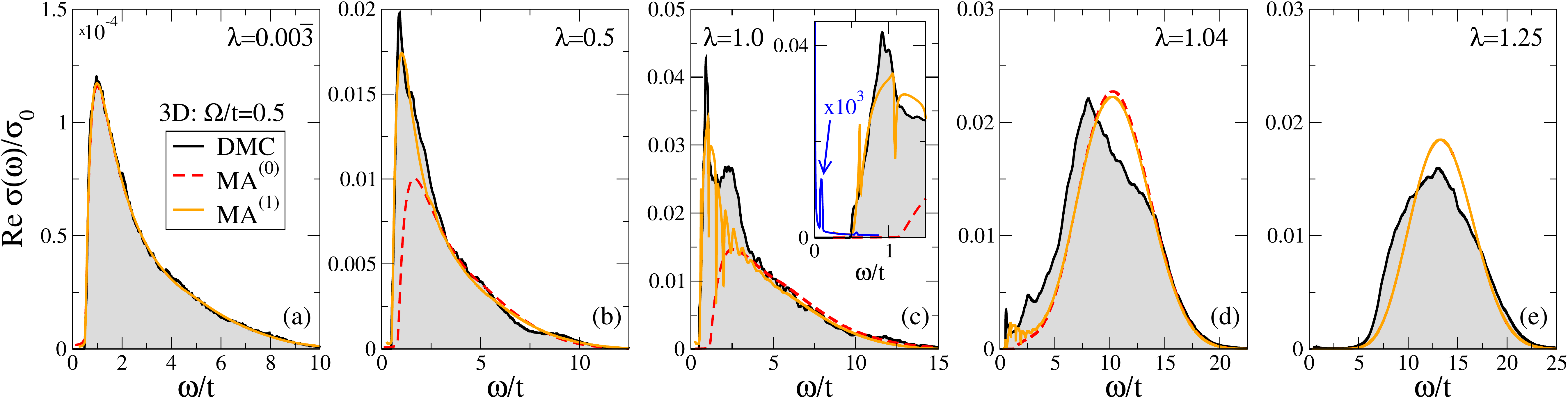}
\caption{(color online) OC in units of $\sigma_0=\pi e^2 t^2$ for a
  3D Holstein polaron, calculated 
  using MA$^{(0)}$, MA$^{(1)}$ and DMC. The inset in panel (c)  shows
   the absorption onset, and the no-vertex correction  OC at $k_BT=0.01\Omega$,
   scaled by $10^3$ (thin blue line).
\label{fig:cond_3D}}
\end{figure*}

In Fig. \ref{fig:cond_3D}, we plot the first, to our knowledge, 
complete set of OC curves reported for a 3D Holstein polaron, using
both MA and DMC methods. On the whole, the agreement is excellent,
especially between MA$^{(1)}$ and DMC. We find that MA$^{(1)}$
captures all the qualitative features of the full OC, as well as being
able to resolve finer structure near the absorption onset. In
particular, the ``shoulder'' that develops on the low-energy end of
the OC spectrum could never be captured with perturbational methods.
In the asymptotic $\lambda \rightarrow 0, \infty$ limits the curves
are nearly indistinguishable, as expected since MA becomes exact in
these limits.  In the crossover regime $\lambda \sim 1$, the
differences between MA$^{(0)}$ and MA$^{(1)}$ are largest, as are
those between MA and DMC results. Nevertheless, MA does a good job
overall, especially considering that it is an efficient analytical
approximation.

The shapes of the OC curves can now be understood using
Eq. (\ref{eq:cond_MA}) and the data shown in Figs. \ref{fig:Pn_3D} and
\ref{fig:fn_3D}. For small $\lambda$, 
$P_{n=1}$ is dominant and the OC is basically proportional to
$f_{n=1}(\omega)$. Indeed, its rich structure is clearly visible in
the OC, as is its threshold $\omega_{\rm th}=\Omega$. The $n\ge 2$
terms serve only to alter the high-energy tail. In
the small polaron limit, however, $P_{n=1} \rightarrow 0$
and the OC is dominated by large  $n$ contributions. Since 
these $f_n(\omega)$ curves have similar shapes and are shifted by
$\Omega$ with respect to one other, the OC mirrors the LF Poissonian
distribution of  $P_n$. The peak location is also in good
agreement with  $2|E_0|=12 \lambda t$, expected as $\lambda \rightarrow
\infty$ \cite{Emin}. This is because when the electron is moved to
a neighboring site by optical absorption, it loses the polaronic
binding energy $E_0$ and it also leaves behind excited phonons with
the same energy.  The structure of our expression (\ref{eq:cond_MA})
proves that not only the peak energy, but the very shape of the OC
curve is determined by the probabilities $P_n^{(i)}$.
In 1D (not shown), the
$f_n(\omega)$ functions are more peaked and individual
contributions can  be seen in the OC, as $\Omega$-spaced kinks. The
agreement with available 1D numerical data is of similar quality with
 that of Fig. \ref{fig:cond_3D}~\cite{goodvin:10}.

Consider now the DMFT-like no-vertex correction approximation, in
which the two-particle Green's function in Eq. (\ref{n1}) is replaced
by a  convolution of polaron spectral weights (for details, see
the supplementary material). The scaled result for $\lambda=1$, $k_B T =0.01\Omega$, is
shown in Fig.~\ref{fig:cond_3D}(c). A prominent feature is the
peak (marked by arrow) below $\omega_{\rm th}$, in agreement with
low-$T$ data 
in Ref. ~\cite{Fratini}, where it was identified as an excitation
from the GS into the first bound-state. Indeed, the peak is at
$\sim0.12t$, their energy difference. The peak likely vanishes at
$T=0, \eta=0$, however this shows that a naive consideration
of the convolution may lead to a qualitatively wrong idea of the OC
structure.  DMC and MA results do not show this peak, although the
second bound state is visible in their spectral
weight (see the supplementary material). Its absence implies a vanishing matrix element
in Eq. (\ref{n2}), likely due to the different symmetry of the polaron
wavefunction in the two states~\cite{bayo}. Clearly,
Eq. (\ref{eq:cond_MA}) properly accounts for such selection rules.
More analysis is provided in the supplementary material found at the end of this article.

\begin{table}[b]
\begin{tabular}{l | c  c  c  c  c}
$\lambda=$ & $0.00\bar{3}$ 
& $\quad$ $0.5$&  $\quad$ $1.0$ & $\quad$
  $1.04$& $\quad$ $1.25$
\\ \hline 
MA$^{(0)}$ & 0.0003 & $\quad$ 0.037 & $\quad$ 0.084 & $\quad$ 0.180 & $\quad$ 0.145 \\ 
MA$^{(1)}$  & 0.0003 & $\quad$ 0.048 & $\quad$ 0.107 & $\quad$ 0.180 & $\quad$ 0.145 \\  
DMC         & 0.0003 & $\quad$ 0.050 & $\quad$ 0.125 & $\quad$ 0.190 & $\quad$ 0.150 \\ 
\end{tabular}
\caption{Total integrated OC $S(\infty)/\sigma_0$  for the parameters used in 
  Fig. \ref{fig:cond_3D}, for MA$^{(0)}$, MA$^{(1)}$, and DMC.
\label{table:fsum_3D}}
\end{table}

To further prove the accuracy of MA, we compare in Table
\ref{table:fsum_3D} the total integrated OC, $S(\infty) =
\int_0^{\infty} d\omega \, \sigma(\omega)$, for  MA and DMC. The
agreement is again very good, particularly at small and 
large $\lambda$, where we typically find less than $5\%$
difference. At intermediate couplings MA accounts for over
$85\%$ of the total integrated OC, which is quite remarkable for an
analytical approximation.

Knowledge of the total integrated OC can also be used to verify that
MA satisfies the \emph{f}-sum rule, given by \cite{maldague:77}:
\begin{equation}
-e^2 a E_{\mr{kin}} = \frac{1}{\pi}{\cal D} + \frac{2}{\pi} S(\infty),
\end{equation}
where $E_{\mr{kin}}$ is the GS polaron kinetic energy and ${\cal
  D}=\pi e^2 a/m^*$ is the Drude weight~\cite{kohn:64}, $m^*$ being
the polaron effective mass. Both these quantities can be calculated
from the GS properties of the polaron, using known MA methods
\cite{berciu:06}. In all cases reported here, we have verified that
the total integrated OC matches the expected value from the
\emph{f}-sum rule to at least three decimal places.


In conclusion, we have obtained the first accurate analytical
non-perturbational expression for the $T=0$ optical conductivity of a
Holstein polaron.  It explains the
shape  of the OC curve by explicitly relating
it to the  statistics of the polaron cloud. We find no absorption
below the threshold at $\Omega$ even if a 
second bound state exists, proving that Eq. (\ref{eq:cond_MA})
accounts for selection rules. Moreover,   this work can be generalized for OC
studies of polaronic systems with momentum-dependent el-ph
coupling, and also disorder, in any
dimension.

{\em Acknowledgments:} Work supported by NSERC and CIfAR (GLG
and MB), and
RFBR 10-02-00047a (ASM).


\section{Supplementary Material}
\subsection{Calculation of the matrix elements within MA$^{(0)}$}

First, we need to calculate the generalized single particle Green's
functions:
\begin{equation}
\EqLabel{1}
F_{nm}(\mb{q},\mb{Q},\omega)= \langle -{\mb q},n| c_{\mb{q}}
\hat{G}(\omega) c_{\mb{Q}}^{\dagger}|-{\mb Q},m\rangle
\end{equation}
where
\begin{equation}
\EqLabel{2}
|-\mb{q},n \rangle = \frac{1}{\sqrt{N}}\sum_i e^{-i\mb{q}
 \cdot \mb{R}_i} (b_i^{\dagger })^n | 0 \rangle
\end{equation}
We use
Dyson's identity $\hat{G}(\omega) =\hat{G}_0(\omega) + \hat{G}(\omega)
V \hat{G}_0(\omega)$ where $V$ is the el-ph interaction and
$\hat{G}_0(\omega)$ is the resolvent for ${\cal H}_0={\cal H}-V$, plus
the MA$^{(0)}$ one-site cloud restriction, to find $F_{nm}(\mb{q}, \mb{Q},
\omega) = G_0(\mb{Q}, \omega - m\Omega) [\delta_{\mb{q},\mb{Q}}
  \delta_{n,m} n!  + mgf_{n,m-1}(\mb{q},\omega) +
  gf_{n,m+1}(\mb{q},\omega) ] $. Here, the free propagator is
$G_0(\mb{k},\omega) = (\omega +i\eta - \varepsilon_{\mb{k}})^{-1}$, and
$f_{n,m}(\mb{q},\omega)=
\frac{1}{N}\sum_{\mb{Q}}F_{nm}(\mb{q},\mb{Q},\omega)$ are partial
momentum averages related to the ones calculated in
 Ref.~\cite{berciu:06alt}. They can therefore be calculated
similarly; however, they depend on $\mb{q}$ only through
$\varepsilon_{\mb{q}}$, therefore they are even functions whose contribution to
$\sigma_+(\omega)$  vanishes after the sum over $\mb{q}$ because of
the odd prefactor $\sin q$ coming from the current operator.  As a
result, within MA$^{(0)}$ we can replace   
$$
F_{nm}(\mb{q}, \mb{Q},
\omega) \rightarrow G_0(\mb{Q}, \omega - m\Omega) \delta_{\mb{q},\mb{Q}}
  \delta_{n,m} n!  $$
in Eq. (\ref{simp}). The delta functions remove the sums over $m, \mb{Q}$. All
that is left, then, is to find the residues $|\langle \psi_0 |
c_{\mb{q}}^{\dagger} | -{\mb 
  q},n\rangle|^2$. To achieve this, consider the generalized
single-electron Green's functions:
\begin{equation}
\EqLabel{3}
F_n(\mb{q},\omega) = \langle 0 | c_{\mb{k}=0}^{} \hat{G}(\omega)
c^{\dagger}_{\mb{q}} | -\mb{q}, n \rangle.
\end{equation}
These are also similar to the single particle Green's functions
calculated in Ref. \cite{berciu:06alt}, and can be calculated in
terms of continued fractions by the same means. Within  MA$^{(0)}$,
they are equal to:
\begin{equation} \label{eq:Fn_inter}
F_n(\mb{q},\omega) = \frac{G_0(\mb{q},\omega -
  n\Omega)}{\bar{g}_0(\omega - n\Omega)} A_{n,1}(\omega) G(\mb{k}=0,\omega). 
\end{equation}
where $G(\mb{k},\omega) = \langle 0 | c_{\mb{k}}^{}
\hat{G}(\omega)c_{\mb{k}}^{\dagger}|0\rangle$ is the usual
single-particle polaron Green's function, and we define $A_{n,k}(\omega) =
A_n(\omega)A_{n-1}(\omega)\cdots A_k(\omega)$ where the continuous
fractions are \cite{berciu:06alt,glen:07}:
\begin{equation}
A_n(\omega) = \frac{ng \bar{g}_0(\omega - n\Omega)}{1-g\bar{g}_0(\omega - n\Omega)A_{n+1}(\omega)}
\end{equation}
and 
\begin{equation}
\bar{g}_0(\omega) = \frac{1}{N} \sum_{\mb{q}} G_0(\mb{q}, \omega)
\end{equation}
is the momentum average of the bare propagator. 

On the other hand, from its Lehmann representation:
\begin{equation}
F_n(\mb{q}, \omega) = \sum_{\alpha} \frac{\langle 0 | c_{\mb{k}=0}^{}
  | \psi_{\alpha}^{} \rangle \langle \psi_{\alpha}^{} |
  c_{\mb{q}}^{\dagger} | -\mb{q},n \rangle}{\omega - E_{\alpha} +
  i\eta}, 
\end{equation}
so its residue for $\omega=E_0$, the GS energy, is the product
$\langle 0 | c_{\mb{k}=0}^{} | \psi_{\alpha}^{} \rangle \langle
\psi_{\alpha}^{} | c_{\mb{q}}^{\dagger} | -\mb{q},n \rangle$. The
first term is related to the GS quasiparticle weight $Z_0$, and the second
is the overlap that we need. It follows that:
\begin{equation} \label{eq:matrix_element}
\langle \psi_0 | c_{\mb{q}}^{\dagger} | -\mb{q}, n \rangle =
\frac{G_0(\mb{q}, \omega-n\Omega)}{\bar{g}_0(\omega - n \Omega)}
\sqrt{Z_0} A_{n, 1}(E_0), 
\end{equation}  
The $\omega$-independent part can be shown to be related to the
probability to have $n$-particles at the electron site in the GS. In
Ref. \cite{berciu:08alt}, these probabilities were found to be:
\begin{equation}
\EqLabel{5}
P_n^{(0)} = {Z_0\over n!} 
\left|A_{n,1}(E_0)\right|^2 
\end{equation}
The $\omega$-dependent terms can be grouped together with the one
from $F_{nm}(\mb{q},\mb{Q},\omega)$ into the corresponding
function $f_n^{(0)}(\omega)$, whose expression is given in Eq. (\ref{fn}).

The MA$^{(1)}$ calculations proceed along similar lines, but are much
more involved because of the enlarged basis. The full details 
will be presented in a longer publication.

\subsection{The approximation of no current vertex corrections}

The essence of this approximation is to replace the two-particle
Green's functions appearing in $\sigma^+(\omega)$, namely $\langle
\psi_0 | c^\dagger_{\mb{Q}} c_{\mb{Q}} \hat{G}(\omega)
c^\dagger_{\mb{q}} c_{\mb{q}}|\psi_0\rangle$, by a convolution of the
single-particle spectral weights. The procedure is reviewed in detail in
Ref. \cite{loos}. The regular part of the conductivity, at finite $T$
and on a cubic lattice,
is then found to be [their Eq. (19)]:
\begin{multline}
\EqLabel{5}
\sigma(\omega) = {4\sigma_0\over \omega  N} \sum_{\mb{k}}^{}
  \sin^2 k \int_{-\infty}^{\infty} d\omega'
  A(\mb{k},\omega')A(\mb{k},\omega+\omega')
 \\ \times \left[f(\omega')-f(\omega+\omega')\right].
\end{multline}
Here $A(\mb{k},\omega)$ is the polaron spectral weight and
$f(\epsilon)=(e^{\beta\epsilon}+1)^{-1}$ is the Dirac function. This
is the same formula used in DMFT, see for example Eq. (3) in
Ref. \cite{Fratini1}. The difference is that in DMFT, the integral over
the Brillouin zone is replaced by an integral over energies $\epsilon =
\epsilon_{\mb{k}}$ (they also replace $\omega' \rightarrow \nu$). This
explains the appearance of the density of 
states $N_\epsilon$ of the free lattice, and of the DMFT vertex
$\Phi_\epsilon$ which is the equivalent of the $\sin^2 k$ in the above
equation. Of course, within DMFT one uses the DMFT self-energy in the
spectral weight.

In order to limit the quantitative differences of the DOS used in DMFT, which
is for a Bethe lattice, whereas our other results are for a cubic
lattice, we use 
Eq. (\ref{5}) for a cubic lattice to test this approximation. Also,
for convenience, 
 we use the MA self-energy in 
the spectral weight. This is simpler because in order
to calculate the DMFT 
self-energy, one needs to 
go through iterations to reach self-consistency. In any event, the
MA and DMFT self-energies are qualitatively and even quantitatively in
fairly good agreement, as shown in the Appendix of
Ref. \cite{berciu:06alt}. Moreover, it has been shown that the MA spectral
weights are very accurate, satisfying eight sum rules exactly  (within
MA$^{(1)}$), therefore using the MA self-energy and its corresponding
spectral weight can lead, at
most, to  modest quantitative differences.

Since the MA$^{(1)}$ self-energy is momentum-independent
(momentum dependence appears only from the MA$^{(2)}$ level), we can
carry the integral over the Brillouin zone as follows. We
rewrite the spectral weights as 
\begin{widetext}
$$A(\mb{k},\omega)=-{1\over \pi}
G(\mb{k},\omega) =  {i\over 2\pi}\left[{1\over \omega+i\eta  -
	\epsilon_{\mb{k}} - 
  \Sigma(\omega)} - {1\over \omega-i\eta  - \epsilon_{\mb{k}} -
  \Sigma^*(\omega)} \right]
$$
The product of two spectral weights will result in a sum of 4 terms
that can be further factorized, for example:
\begin{multline}
{1\over \omega'+i\eta  -	\epsilon_{\mb{k}} -   \Sigma(\omega')}\cdot
{1\over \omega+\omega'+i\eta  -	\epsilon_{\mb{k}} -   \Sigma(\omega+\omega')}
\\
= 
{1\over \omega - \Sigma(\omega+\omega')+\Sigma(\omega')}
\left[{1\over \omega'+i\eta  -	\epsilon_{\mb{k}} -   \Sigma(\omega')}-
{1\over \omega+\omega'+i\eta  -	\epsilon_{\mb{k}} -
  \Sigma(\omega+\omega')}\right\} 
\end{multline}
\end{widetext}
As a result, the integral over the Brillouin zone now becomes a simple
momentum average of the bare propagator (with shifted frequency) times
the $\sin^2{k_x}$ term. One integral can be done analytically, and the
other two numerically like other similar momentum
averages calculated for  MA$^{(2)}$, see Ref.  \cite{berciu:06alt}. One
is then left with evaluating the integral over $\omega'$ numerically as well.

One  can approach the $T=0$ limit either directly by replacing the
Fermi-Dirac distribution with a Heaviside function, as done in
Ref. \cite{loos}; or by calculating the $T\rightarrow 0$ limit of the
ratio detailed in Refs.  \cite{Fratini1,Fratini2}. Either approach has
its own numerical challenges, and overall we find both procedures to
be more time consuming than evaluating our OC formula, even when the
computationally trivial MA self-energy is used in the spectral
weight. Note, also, that our 
OC formula is formulated directly for $T=0$ and therefore one needs not
worry about how to properly approach this limit.

\begin{figure}[b]
\includegraphics[width=\columnwidth]{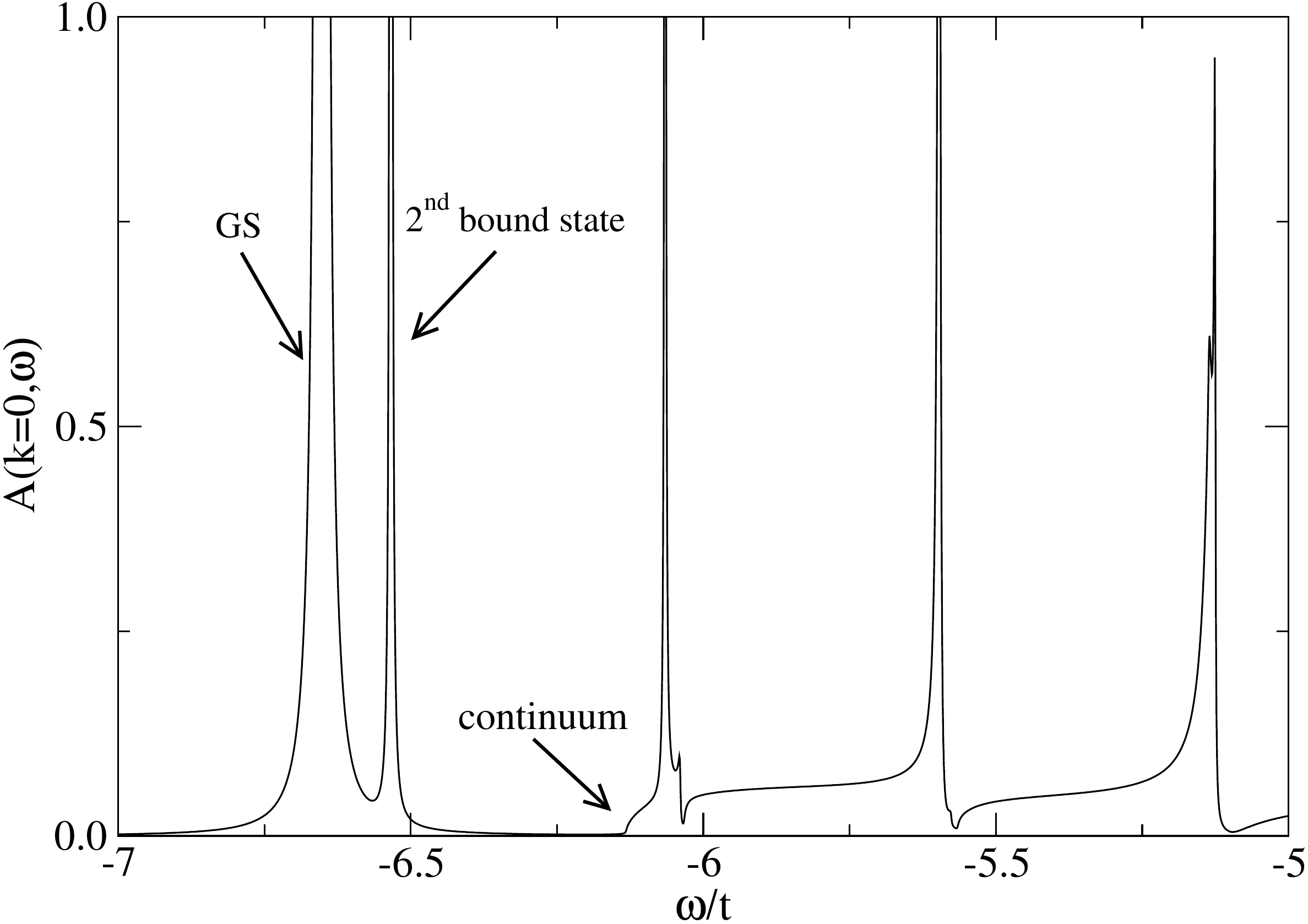}
\caption{\label{supp1} Spectral weight $A(\mb{k}=0,\omega)$ for the 3D
  Holstein polaron, for $\lambda=1, \Omega=0.5t, \eta =10^{-3}$. The
  GS energy is $E_0= -6.65 t$, and the first bound state is at
  $E_1=-6.53t$. The continuum onset is visible at $E_0+\Omega$.}
\end{figure}

The $\mb{k}=0$ spectral weight for the polaron, at intermediary
coupling $\lambda=1$, is shown in
Fig. \ref{supp1}. The GS peak still 
has considerable weight $Z_0\sim 0.5$, and just above it we see the peak
corresponding to the second bound state, followed by the continuum and
higher energy features. The rough expectation is that a  convolution
of this curve with itself will 
show a first peak when the two curves are shifted by $
E_1-E_0$,  the energy difference between the two bound
states; also, absorption into the continuum will appear for frequencies
$\omega > \Omega$.

Thus, the naive expectation based on the convolution  is that as
soon as  the second bound state is formed, there should be absorption into it, at
energies below the threshold $\Omega$. This is also what is
shown by the full DMFT calculations at very low temperatures, see
curve of Fig. 2 and its interpretation from Fig. 3 in
Ref. \cite{Fratini1}, in qualitative agreement with our own
calculation for  $k_BT = 0.01 \Omega$, shown in Fig. 3(c) of the main
article. 

In reality, very careful analysis suggests that the weight of this peak
decreases monotonically with $T$, and therefore at $T=0$ it is very
likely that this approach also predicts no absorption below the
threshold at $\Omega$, in agreement with the DMC and MA
approaches. Our understanding of the disappearance of this peak is
that it comes from the fact that the $k=0$ convolution is multiplied by
$\sin^2 k$, so strictly speaking there is no contribution to the OC
from the GS momentum, unless some ``thermal broadening'' is allowed at
non-zero temperatures.

Since the exact DMC calculation finds no such
absorption into the second bound state at any $\lambda$, the matrix
element for this transition 
must be zero, likely because of the different symmetry of the phonon
clouds in the two states, as discussed in detail for 1D systems in
Ref. \cite{new}. Clearly, our MA-based OC formula captures this
selection rule properly -- note that the generalized
single-particle Green's functions that appear in it are different from the
Green's function that gives the spectral weight. 

On the other hand, there is nothing in the spectral weight that
encodes the symmetry of the wavefunction associated with a bound
state, for example one cannot distinguish the parity of a state by
looking at the plot of the spectral weight. Thus, it seems to us that
generically, one should  expect the convolution approximation to fail
to properly account for selection rules.
 
In fact, this statement can be tested in 1D, where as shown in
Ref. \cite{new}, one may expect higher energy bound states, some of
which are optically active. In other models, it has been suggested
that there may even be bound-states below the continuum which are not
visible in the spectral weight, because their quasiparticle weight
vanishes for symmetry reasons \cite{bayoalt} -- if such states are
optically active, absorption associated with them would certainly not
be predicted by the convolution approximation.  Such a study will be
undertaken elsewhere, and should clarify whether the agreement in 3D
regarding the lack of sub-threshold absorption in the no-current
vertex approximation is accidental, or it actually has a more robust
explanation. In any event, this is probably somewhat of an academic
discussion, since such multiple bound states appear only at large
effective couplings $\lambda$ where the bulk of the optical absorption
is at very much higher energies, and in measurements the very small signal at the
threshold may be buried in noise.

Be that as it may, to us it appears that besides being numerically
more difficult to evaluate and less tested in terms of its accuracy
for sum rules etc, the no-current vertex approximation also fails to
provide the insight into the overall evolution of the OC curves that
is afforded by our new formula. Moreover, MA has already been
demonstrated to generalize very successfully to models where the
coupling depends on the phonon and electron momentum, and we expect
similar success in modeling their OC.  Such models
have strongly momentum-dependent self-energies, and are unlikely to be
described accurately with a local-approximation like DMFT.

\end{document}